\documentclass[epj]{webofc}
\usepackage[utf8]{inputenc}
\usepackage[varg]{txfonts}
\usepackage{booktabs}
\usepackage{xcolor}
\definecolor{darkred}{rgb}{0.4,0.0,0.0}
\definecolor{darkgreen}{rgb}{0.0,0.4,0.0}
\definecolor{darkblue}{rgb}{0.0,0.0,0.4}
\usepackage[bookmarks,linktocpage,colorlinks,
    linkcolor = darkred,
    urlcolor  = darkblue,
    citecolor = darkgreen]{hyperref}
\usepackage{subfigure}

\wocname{EPJ Web of Conferences}
\woctitle{Lattice2017}

\begin{document}
%%%%%%%%%%%%%%%%%%%%%%%%%%%%%%%%%%%%%%%%%%%%%%%%%%%%%%%%%%%%%%%%%%%%%%%%%%%%%

\selectlanguage{english}

\title{Influence of magnetic fields on the color screening masses}

\author{
  \firstname{Claudio}   \lastname{Bonati}     \inst{1}
  \fnsep \thanks{Present address:   Dipartimento di Fisica dell’Università di Pisa and
  INFN Sez. Pisa, Largo Pontecorvo 3, 56127 Pisa, Italy.} \and
  \firstname{Massimo}   \lastname{D'Elia}     \inst{2} \and
  \firstname{Michele}   \lastname{Mesiti}     \inst{2} \and
  \firstname{Francesco} \lastname{Negro}      \inst{2} \and
  \firstname{Andrea}    \lastname{Rucci}      \inst{2} \fnsep \thanks{Speaker, \email{andrea.rucci@pi.infn.it}} \and
  \firstname{Francesco} \lastname{Sanfilippo} \inst{3}
}

\institute{%
  Dipartimento di Fisica e Astronomia dell’Università di Firenze and
  INFN Sez. Firenze, Via Sansone 1, 50019 Sesto Fiorentino
  (FI), Italy
  \and
  Dipartimento di Fisica dell’Università di Pisa and
  INFN Sez. Pisa, Largo Pontecorvo 3, 56127 Pisa, Italy.
  \and
  INFN Sez. Roma Tre, Via della Vasca Navale 84, 00146
  Rome, Italy.
}

\abstract{%
  We present some recent results obtained in the study of the color
  magnetic and electric screening masses in the QCD plasma. In
  particular, we discuss how the masses get modified by strong
  external fields which are expected to be created in physical
  situations such as heavy-ion collisions.}

\maketitle

\section{Introduction}\label{intro}

Strong magnetic fields of the order of the QCD scale can be found in
many physical situations and phenomena such as the early universe
\cite{Vachaspati:1991nm, Grasso:2000wj} or non-central heavy ion
collisions \cite{Skokov:2009qp,Tuchin:2013ie}. In these contexts,
fields with intensities up to $10^{16}~\rm{Tesla}$
($|e|B\sim1~\rm{GeV}^2$) are expected to be produced. They may
influence the properties of strongly interacting matter and many
theoretical studies have been devoted to this argument (see reviews in
Refs.~\cite{Kharzeev:2012ph,Miransky:2015ava}). As regards the color
interaction, in several studies the effects of a magnetic background
on the static quark-antiquark potential has been investigated.
\cite{Rougemont:2014efa,Ferrer:2014qka,Chernodub:2014uua}.  In recent
lattice studies \cite{Bonati:2014ksa,Bonati:2016kxj} it has been shown
that the potential gets modified and relevant consequences may arise
at the level of heavy meson production and spectrum (see
\cite{Bonati:2015dka,Bonati:2017uvz} and the references herein). In
particular, at zero temperature the interaction becomes anisotropic
due to the string tension $\sigma$ which decreases in the direction
parallel to the external field $\mathbf{B}$, while it grows on the
orthogonal plane; at larger $T$, in the confined regime below the
pseudo-critical temperature $T_c$, the external field is responsible
of a precocious loss of the deconfining properties, in agreement with
the picture of a decreasing $T_c$ due to the field itself.

In our work, we investigated the properties of the heavy
quark-antiquark interaction in the deconfined region above $T_c$
\cite{Bonati:2017uvz}. In the quark-gluon plasma, the color interaction
gets screened by the thermal medium and two different screening
lengths (or masses) can be defined corresponding to the contribution of
color-electric and color-magnetic gluons. As a consequence, the
production rate of heavy quark bound states is expected to be
suppressed. As argued in the seminal paper in Ref.~\cite{Matsui:1986dk},
the bound state formation gets hindered by the shortening of the
screening length when it becomes comparable to the mean radius of the
state itself. Our effort has been devoted to the study, on the
lattice, of the possible effects of a strong magnetic background on
the screening lengths.

This document is organized as follows. In Section \ref{setup} we describe
the recipe we followed to define and extract the screening masses and
the numerical setup. In Section \ref{results} we show and discuss our
results. Then, we will draw our conclusions in Section \ref{conclusions}.

\section{Setup}\label{setup}

\subsection{Definitions}

It is known that a perturbative definition of the screening masses in
QCD, based on the study of the pole structure of finite temperature
gluon propagators, presents difficulties. With this approach, an
expression for the color-electric mass can be found at leading order,
but the computation turns out to get into troubles at higher order due
to divergences of the obtained expressions
\cite{Gross:1980br,Nadkarni:1986cz,Braaten:1994pk}. This
problem has been overcome and it has been shown that a
non-perturbative definition of the screening masses can be obtained
\cite{Nadkarni:1986cz,Braaten:1994qxx,Arnold:1995bh} by studying the
large distance behaviour of suitable gauge-invariant correlators. The
correlator between Polyakov loop is traditionally used
\begin{equation}\label{eq:polycorr}
  C_{LL^\dagger}(\mathbf{r},T) =
  \left\langle\mathrm{Tr}L(\mathbf{0})\mathrm{Tr}L^{\dagger}(\mathbf{r})\right\rangle ~,
\end{equation}
where
\begin{equation}\label{eq:polyloop}
  L(\mathbf{r}) =
  \frac{1}{N_c}\mathcal{P}\exp\left(-ig\int_{0}^{1/T}A_{0}(\mathbf{x},\tau)d\tau\right)
\end{equation}
is the Polyakov loop operator, $\mathcal{P}$ is the path-order
operator and $N_c$ is the number of colors. The correlator
$C_{LL^{\dagger}}$ has been largely investigated on the lattice due to
its relation \cite{McLerran:1981pb} with the free energy
$F_{Q\bar{Q}}(\mathbf{r},T)$ of a static quark-antiquark pair
\begin{equation}\label{eq:polycorrfreeenergy}
  F_{Q\bar{Q}}(\mathbf{r},T) = -T\log C_{LL^\dagger}(\mathbf{r},T) ~.
\end{equation}
Therefore, this observable is the finite temperature counterpart of
the Wilson loop which is commonly used to extract the static potential
in systems at zero temperature. In our analysis we also made use of
the correlator $C_{LL}$ whose definition retraces the one above,
\begin{equation}\label{eq:polycorr}
  C_{LL}(\mathbf{r},T) =
  \Big\langle\mathrm{Tr}L(\mathbf{0})\mathrm{Tr}L(\mathbf{r})\Big\rangle ~,
\end{equation}
and whose large distance behaviour turns out to be substantially the
same of $\rm{Re} C_{LL^{\dagger}}(\mathbf{r},T)$ in the deconfined
regime \cite{Bonati:2017uvz}.

In order to extract the color-electric and color-magnetic screening
masses, symmetries can be used to separate the two
contributions. Under Euclidean time-reversal
$\mathcal{R}: \tau\to-\tau$ the gluon vector and time components
$A_{i}(\mathbf{x},\tau)$ and $A_{0}(\mathbf{x},\tau)$ are even and
odd, respectively. Using this property it is straightforward to show
that $\mathcal{R}: L\to L^{\dagger}$ and hence one can define the
following combinations of Polyakov loops
\begin{equation}
  L_{M} = \frac{1}{2}\left(L+L^{\dagger}\right) \qquad
  L_{E} = \frac{1}{2}\left(L-L^{\dagger}\right)
\end{equation}
which belong to the magnetic and electric sector, separately. These
object can be further decomposed by using the charge conjugation
operator $\mathcal{C}$ which acts on the Polyakov loop as
$\mathcal{C}: L\to L^{*}$. In this way we can write
\begin{equation}
  L_{M^{\pm}} = \frac{1}{2}\Big(L_{M} \pm L_{M}^{*}\Big) \qquad
  L_{E^{\pm}} = \frac{1}{2}\Big(L_{E} \pm L_{E}^{*}\Big) ~,
\end{equation}
where the $\mathcal{C}$ eigenvalues are indicated by the substripts
$\pm$. From the expression above we find that
$\rm{Tr}L_{M^{-}}=\rm{Tr}L_{E^{+}}=0$, meaning that there is no
overlap with the magnetic odd and the electric even sectors. Magnetic
and electric correlator can now be defined as
\begin{equation}\label{eq:magneleccorr}
  C_{M^{+}}(\mathbf{r},T) =
  \Big\langle\mathrm{Tr}L_{M^{+}}(\mathbf{0})\mathrm{Tr}L_{M^{+}}(\mathbf{r})\Big\rangle
  - |\langle\mathrm{Tr}L\rangle|^2
  \qquad
  C_{E^{-}}(\mathbf{r},T) =
  - \Big\langle\mathrm{Tr}L_{E^{-}}(\mathbf{0})\mathrm{Tr}L_{E^{-}}(\mathbf{r})\Big\rangle ~,
\end{equation}
where in the second definition the sign is conventional and the
disconnected term is not present due to the charge conjugation
symmetry. This decomposition has been introduced in
Ref.~\cite{Arnold:1995bh} and the correlators above have been studied
recently in some lattice studies
\cite{Maezawa:2010vj,Borsanyi:2015yka}. In our case, we accessed these
objects in terms of the correlators $C_{LL^{\dagger}}$ and $C_{LL}$
which are related to the color-magnetic and color-electric by the relations
\begin{equation}
  C_{M^{+}} = +\frac{1}{2}\mathrm{Re}\Big[C_{LL}+C_{LL^{\dagger}}\Big] - |\langle\mathrm{Tr}L\rangle|^2
  \qquad
  C_{E^{-}} = -\frac{1}{2}\mathrm{Re}\Big[C_{LL}-C_{LL^{\dagger}}\Big]  ~.
\end{equation}

Finally, from the large distance behaviour of these correlators it is
possible to extract the screening masses. Indeed, at very high
temperatures it is expected \cite{Braaten:1994qxx,Arnold:1995bh} that
\begin{equation}\label{eq:fitmodel}
  C_{M^{+}}(\mathbf{r},T)\Big|_{r\to\infty} \sim \frac{e^{-m_{M}(T)r}}{r}
  \qquad
  C_{E^{-}}(\mathbf{r},T)\Big|_{r\to\infty} \sim \frac{e^{-m_{E}(T)r}}{r} ~,
\end{equation}
where $m_E$ and $m_M$ are, respectively, the electric and magnetic
masses.

\subsection{Numerical setup}

We used stout smeared staggered fermions (with $N_f=2+1$ at physical
point) and a Symanzik tree-level improved gauge action
\cite{Weisz:1982zw,Curci:1983an}. The partition function in the
presence of an external magnetic field reads as
\begin{equation}\label{partfunc}
  Z(B) = \int\mathcal{D}U e^{-\mathcal{S}_{YM}}
  \prod_{f=u,d,s} \det{({D^{f}_{\rm{st}}[B]})^{1/4}} ~,
\end{equation}
with $\mathcal{D}U$ integration measure of the SU(3) gauge links and
$S_{YM}$ is the tree-level improved gauge action
\begin{equation}
  S_{YM} = -\frac{\beta}{3}\sum_{i;\mu\neq\nu}
  \Bigg(\frac{5}{6}W_{i;\mu\nu}^{1\times1}
  -\frac{1}{12}W_{i;\mu\nu}^{1\times2}\Bigg) ~,
\end{equation}
where the $W$s are the real parts of the trace of square and
rectangular loops. The fermion matrix is
\begin{equation}
  (D^f_{\textnormal{st}})_{i,j} =\  am_f
  \delta_{i,j}+\sum_{\nu=1}^{4}\frac{\eta_{i;\nu}}{2}
  \left(u^f_{i;\nu}U^{(2)}_{i;\nu}\delta_{i,j-\hat{\nu}}
  \right. -\left. u^{f*}_{i-\hat\nu;\nu}U^{(2)\dagger}_{i-\hat\nu;\nu}\delta_{i,j+\hat\nu}
  \right) ~,
\end{equation}
where $U^{(2)}$ are two times stout-smeared gauge links with isotropic
smearing parameter $\rho=0.15$ \cite{Morningstar:2003gk} and the $u$s
are the abelian parallel transports representing the external magnetic
field. We considered a constant and uniform magnetic field pointing
along the $\hat{z}$ direction. In this case, a possible choice of the
U(1) gauge links is
\begin{equation}
  u_{i;y}^{f} = e^{ia^2q_fB_zi_x}~, \quad u_{i;x}^{f}\Big|_{i_x=L_x} = e^{-ia^2q_fL_xi_y}~,
\end{equation}
with the remaining links set to the identity, where $q_f$ is the fermion
charge. Notice that, on a lattice with periodic boundary conditions,
the magnetic field must verify the quantization condition
$|e|B_z = 6\pi b_z / (a^2N_xN_y)$ where $b_z\in\mathbb{Z}$.

In our Monte-Carlo simulations we considered bare parameters given by
$\beta=3.85$, $m_s/m_l = 28.15$ and $am_s=0.0394$ corresponding to a
lattice spacing $a\simeq0.0989~\rm{fm}$, corresponding to physical
pion mass \cite{Aoki:2009sc,Borsanyi:2010cj,Borsanyi:2013bia}. We
used lattices with volumes $48^3\times N_t$ with $N_t=6,8,10$ which
correspond to a physical size of about $5~\rm{fm}$ and temperatures
$T\simeq330~\rm{MeV},250~\rm{MeV},200~\rm{MeV}$ respectively. For each
system we collected statistics of $\sim5\times10^3$ configurations
separated by five trajectories of molecular dynamics. Statistical
noise have been reduced applying a step of HYP smearing
\cite{Hasenfratz:2001hp,DellaMorte:2005nwx}.

For $B=0$ correlators have been computed by averaging over all lattice
directions, while in the presence of the external field along $\hat{z}$
we measured separately the correlators in the $xy$ plane and those
along the $z$ axis, i.e. the orthogonal plane and the parallel axis
with respect to the magnetic field $\mathbf{B}$.

\section{Results}\label{results}

\begin{figure}[thb]
  \centering
  \includegraphics[width=0.45\textwidth,clip]{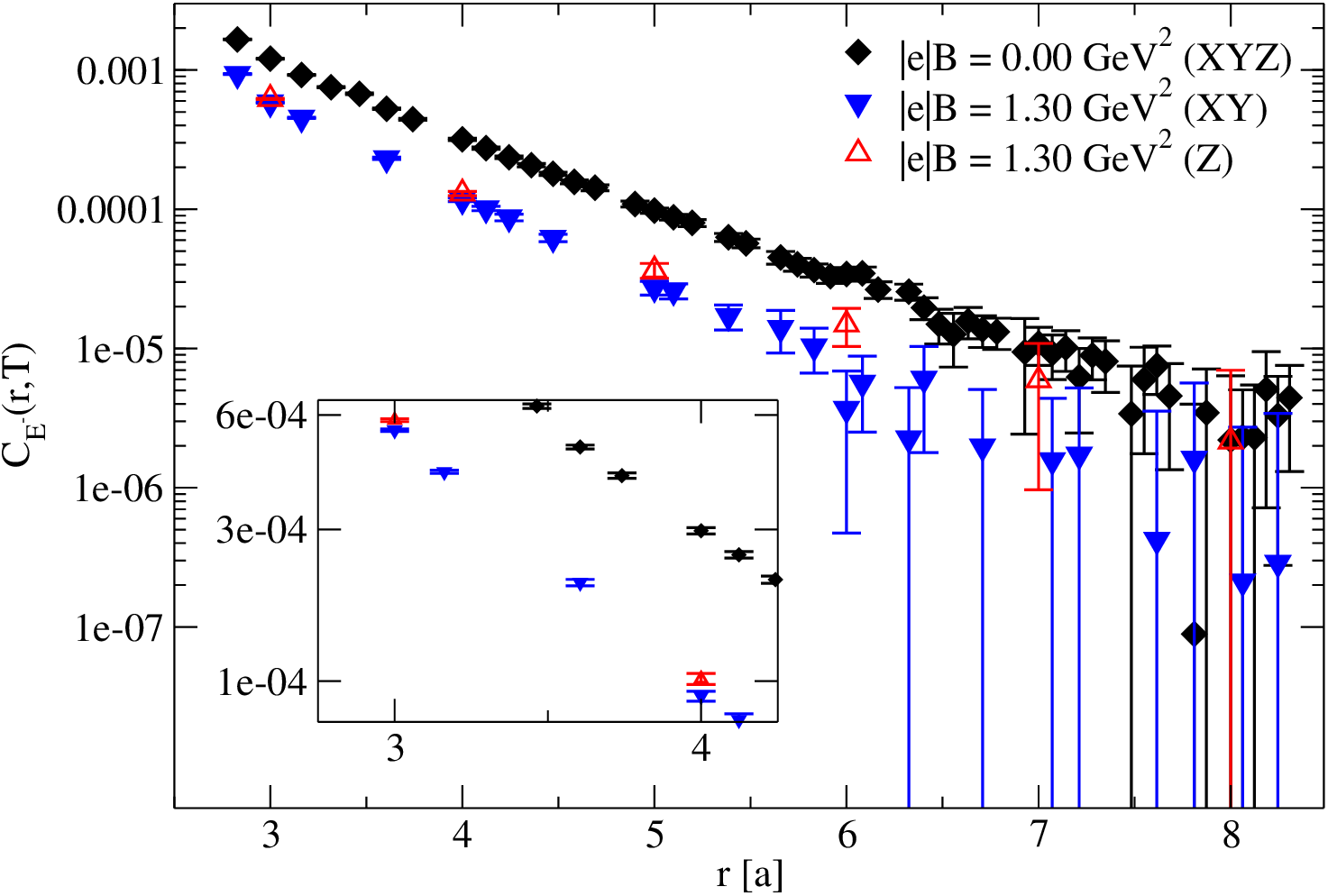}~
  \includegraphics[width=0.45\textwidth,clip]{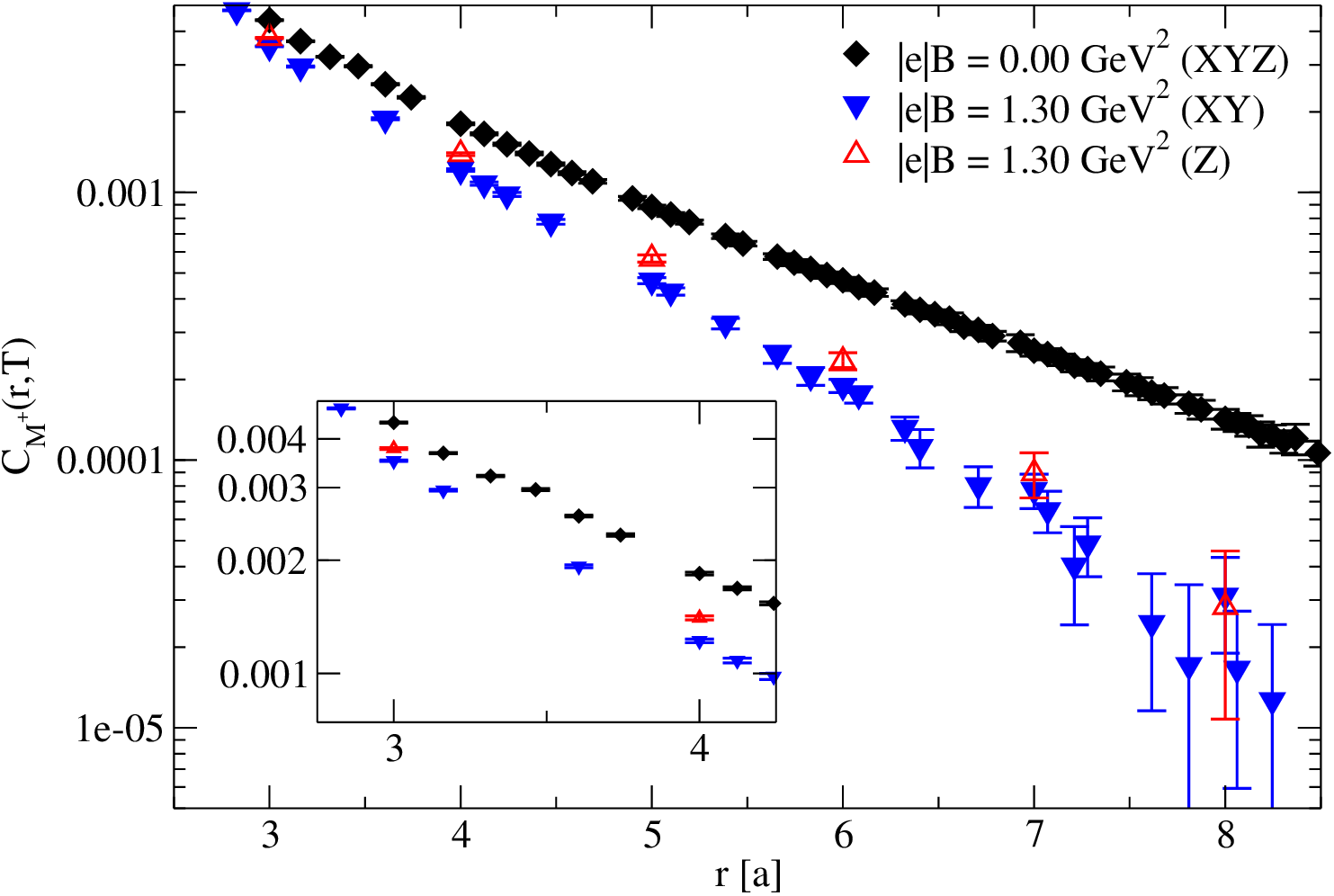}
  \caption{Comparison of both the color-electric $C_{E^{-}}(r,T)$
    (left) and color-magnetic $C_{M^{+}}(r,T)$ (right) correlators
    without magnetic field and with $|e|B\simeq1.30~\rm{GeV}^2$ at our
    lowest temperature $T\simeq200~\rm{MeV}$. At $B=0$ data have been
    obtained averaging over all the directions, while in the presence
    of the external field we separated the contributions along the
    magnetic field (Z) and on the orthogonal plane (XY).}
  \label{fig:magneleccorr}
\end{figure}

Color-magnetic and color-electric correlators as defined in
Eq.~\eqref{eq:magneleccorr} have been computed for several
temperatures and magnetic field intensities. An example of the results
we obtained is shown in Fig.~\ref{fig:magneleccorr}. As can be seen,
when the magnetic field is turned on both the correlators decay
faster, suggesting that the associated screening masses are
increased. Moreover, a slight anisotropy emerge in the correlators and
it seems more pronouced in the magnetic case where the signal is
larger and less noisy with respect the electric one. As regards the
effect of the temperature, our data suggest that the magnetic effects
are reduced when the $T$ grows \cite{Bonati:2017uvz}.

\begin{figure}[thb]
  \centering
  \includegraphics[width=0.45\textwidth,clip]{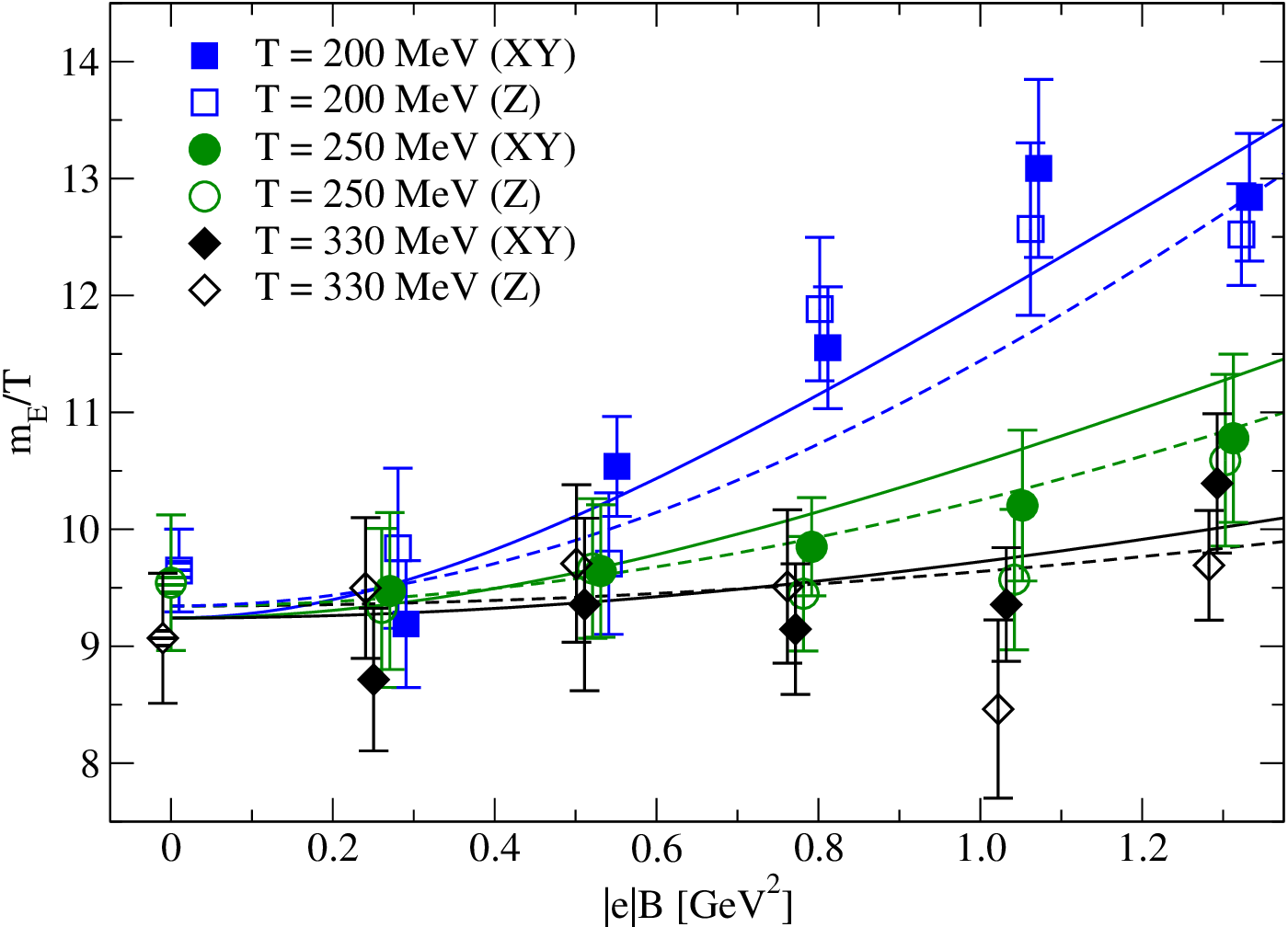}~
  \includegraphics[width=0.45\textwidth,clip]{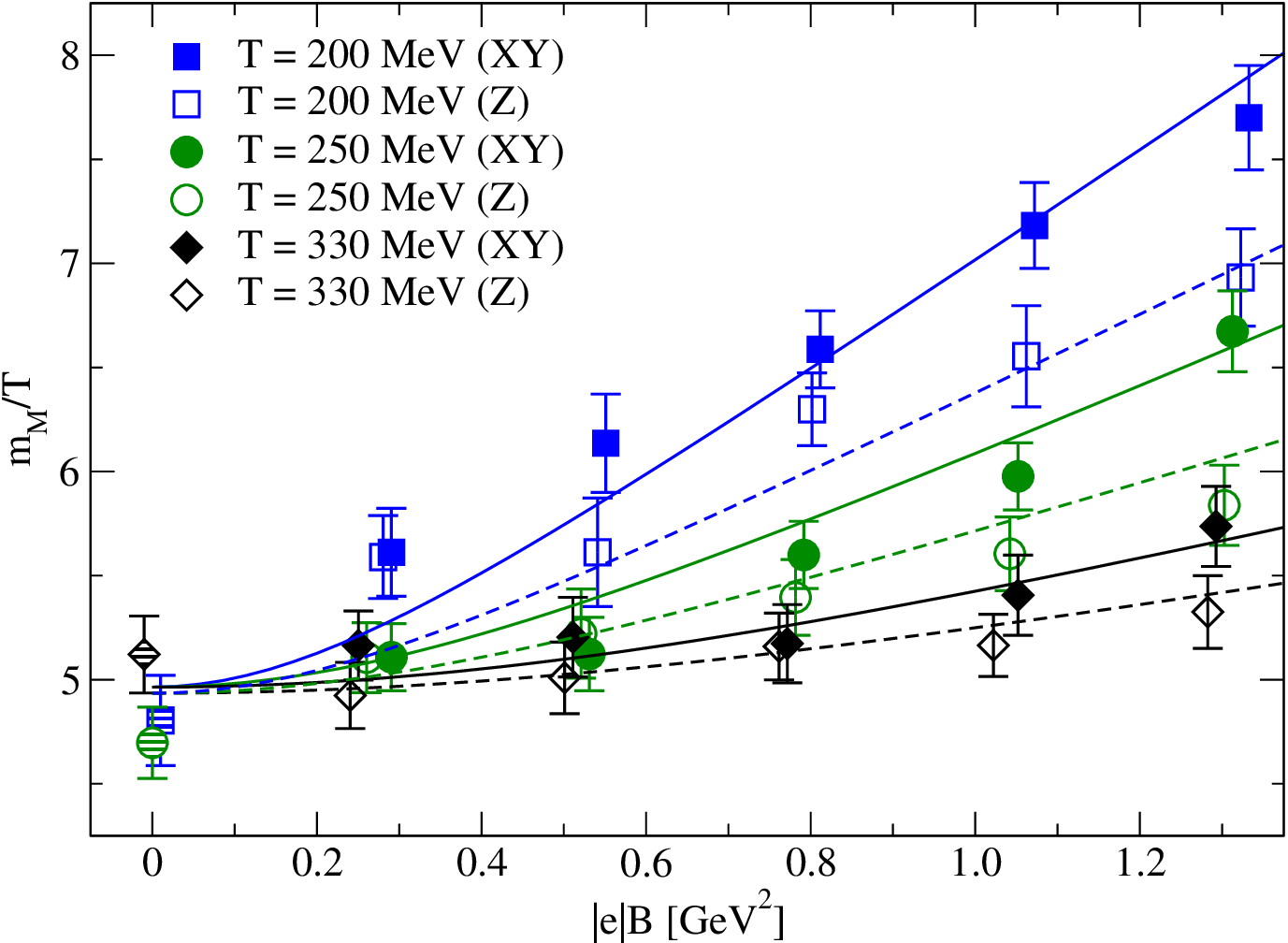}\\~\\
  \includegraphics[width=0.45\textwidth,clip]{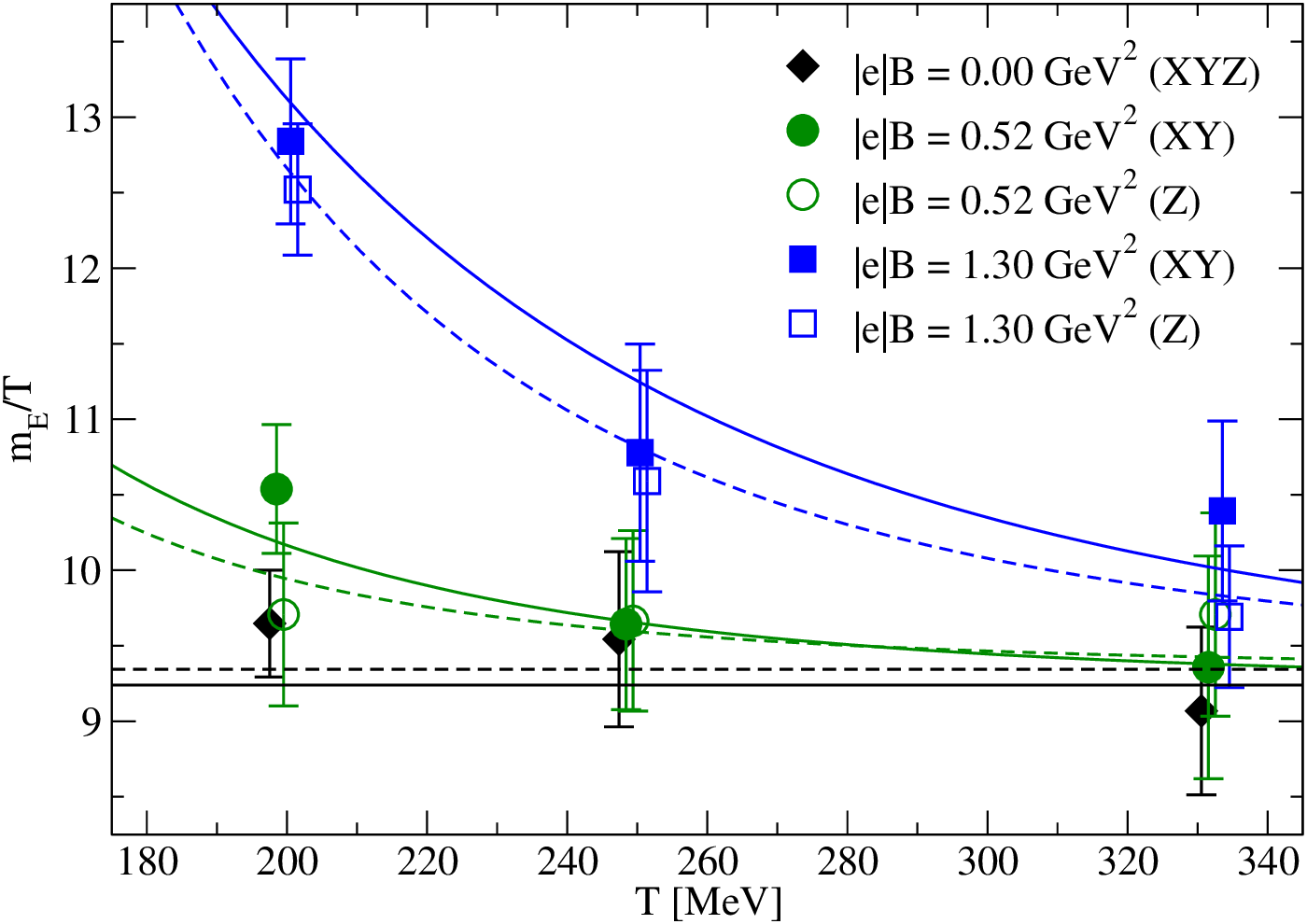}~
  \includegraphics[width=0.45\textwidth,clip]{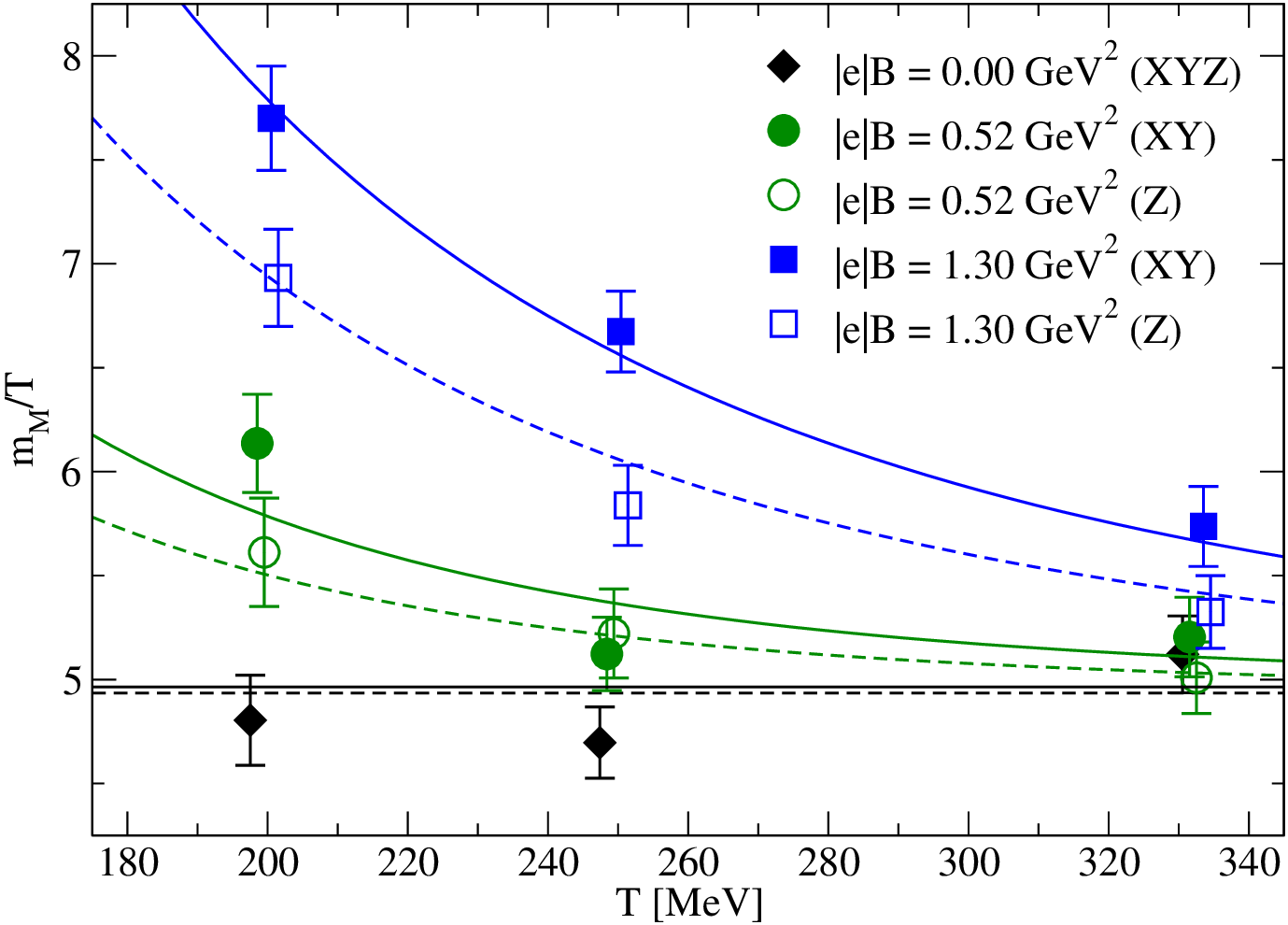}\\~\\
  \includegraphics[width=0.45\textwidth,clip]{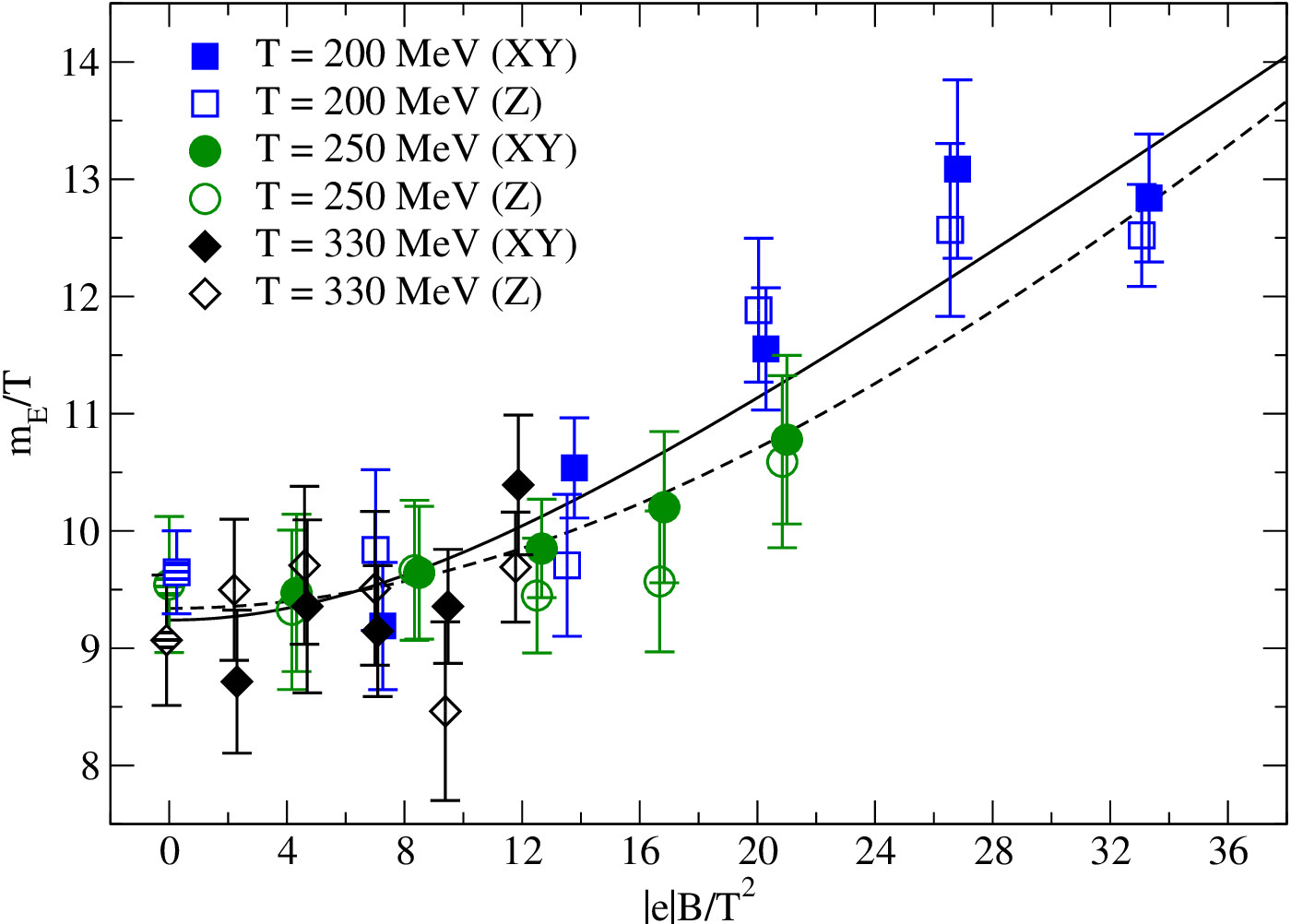}~
  \includegraphics[width=0.45\textwidth,clip]{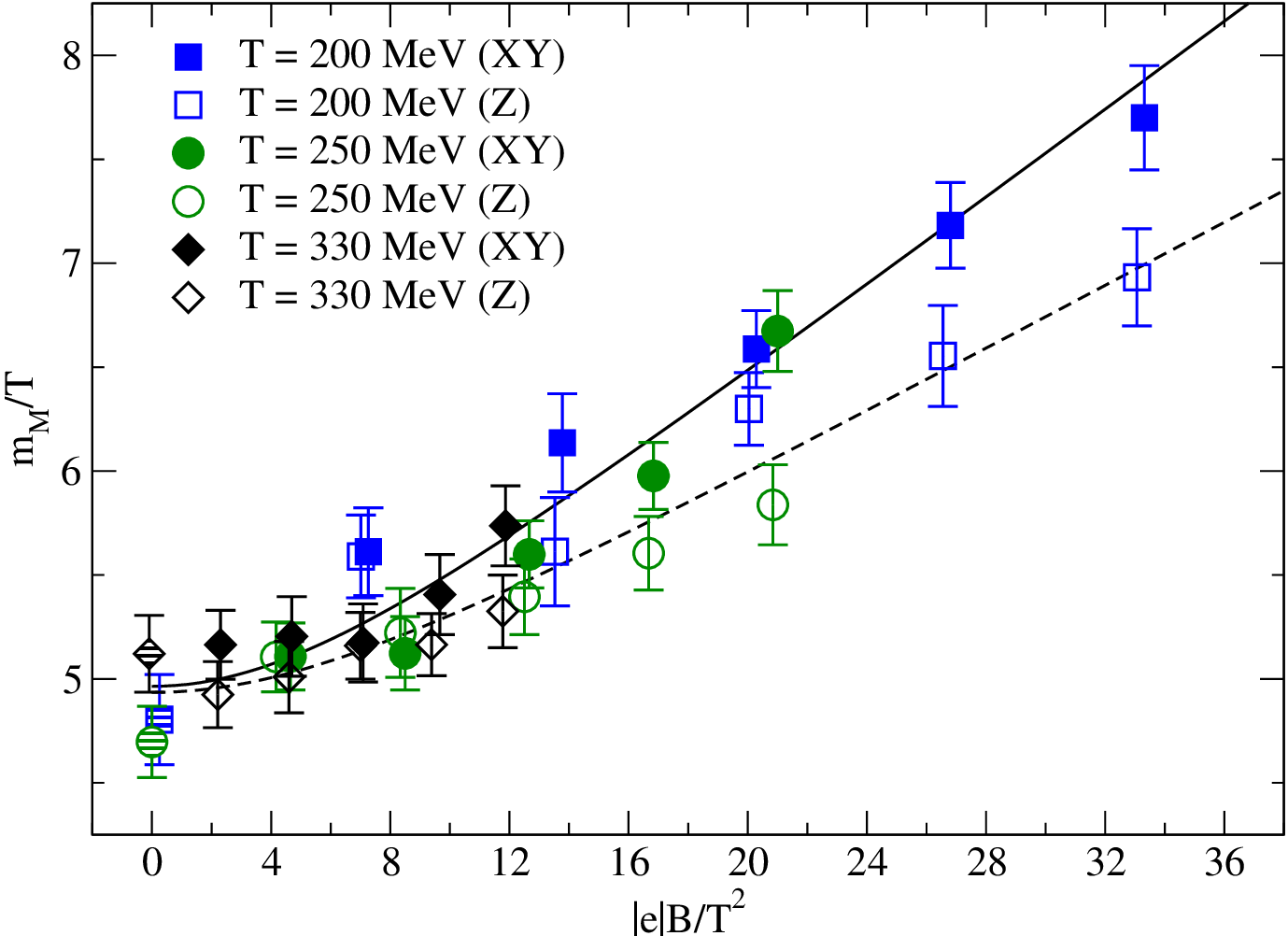}
  \caption{Behaviour of the ratios $m_E/T$ (left) and $m_M/T$ (right)
    as a function of the magnetic field $|e|B$ (up), the temperature
    $T$ (center) and against the dimensionless ratio $|e|B/T^2$
    (down). In all cases data are shown separately for the direction
    parallel (Z) and orthogonal (XY) to the magnetic field. Curves
    associated to the data points are obtained by a regression with
    the model in Eq.~\eqref{eq:ansatz} with the parameters reported in
    Tab.~\ref{tab:fitparams}.}
  \label{fig:masses}
\end{figure}

\begin{figure}[t]
  \centering
  \includegraphics[width=0.45\textwidth,clip]{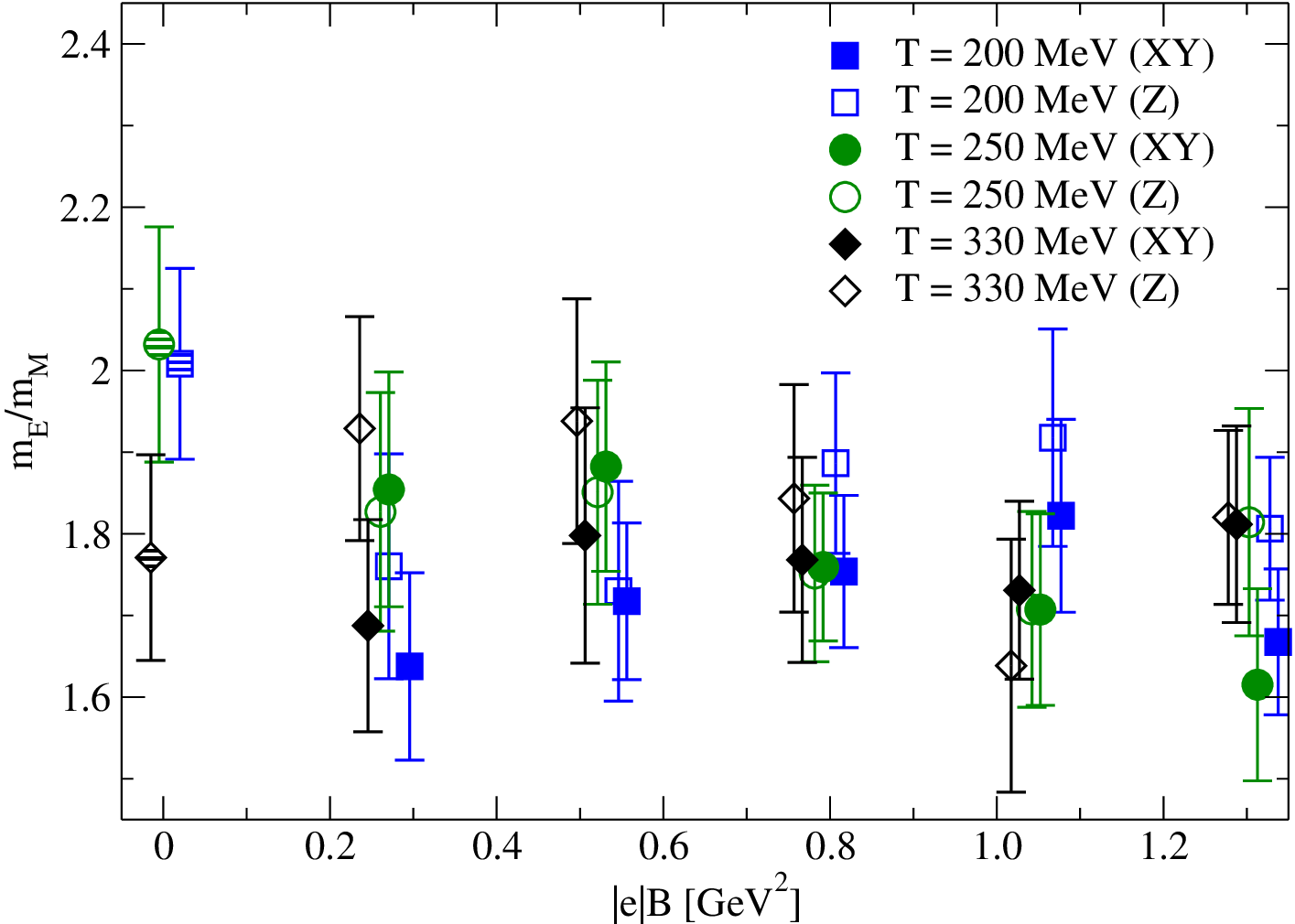}
  \caption{Ratio $m_E/m_M$ of the screening masses as a function of
    $|e|B$ for several different temperatures and separating data
    along the direction parallel and orthogonal to the magnetic
    field.}
  \label{fig:massesratio}
\end{figure}

Our data have been fitted using the model in
Eq.~\eqref{eq:fitmodel}. In order to take into accout correlation
between data, a boostrap resamplig approach has been applied. The
stability of the regression results has been checked by considering
several fit intervals which allowed us to give an estimate of the
systematic uncertainties associated to the procedure. The results we
obtained are shown in Fig.~\ref{fig:masses}. At $B=0$ our data agree
with the expected behaviour, the masses growing linearly with the
temperature and keeping the correct hierarchy $m_E>m_M$. Our findings
are also in accordance with the results obtained on the lattice in
Ref.~\cite{Borsanyi:2015yka} with the same discretization adopted in
our work. In the presence of an external magneti field, our data
suggest that the screening masses grow as a function of $|e|B$,
according to the observations pointed out previously by looking at the
behaviour of the magnetic and electric correlators. In both cases,
$m_E$ and $m_M$ turns out to increase roughly linearly with similar
slope. This behaviour can be noticed also by looking at the ratio
$m_E(T,B)/m_M(T,B)$ in Fig.~\ref{fig:massesratio}: it is essentially
constant in the regime of magnetic field we explored, meaning also
that $|e|B$ do not seems to alter the mass hierarchy. As also guessed
above, the magnetic mass $m_M$ shows an anisotropic behaviour, its
value on the plane orthogonal to the external field being larger than
that on the parallel direction. Conversely, in the case of the
electric mass this effect is not observed, but the effect could be
hindered by the larger noise. In all cases, as can be seen by looking
at the masses plotted against $T$ (see Fig.~\ref{fig:masses}), all the
magnetic effects seem to reduce when the temperature grows.

We tried to find a model describing the observed behaviour of our
data. In order to do so, we looked at a functional form in the
variables $T$ and $|e|B$ retracing the main properties shown by the
data: at $B=0$ the ratios $m_M/T$ and $m_E/T$ are essentially constant
with respect to the temperature; at large $B$ both masses grow almost
linearly. In addition to these, it can be also seen that both masses
seem to be sensitive only to the dimensionless ratio $|e|B/T^2$. The
goodness of this hypothesis can be checked by looking at the behaviour
of the masses in Fig.~\ref{fig:masses}: both $m_E$ and $m_M$
essentially lie on a single curve, sharing the same shape for each
temperature. Finally, it is reasonable to add the requirement that the
functional form must be an analytic function of $B$, so that at small
magnetic fields the dependence should be quadratic. A possible ansatz
that fullfill all these requirements is
\begin{equation}\label{eq:ansatz}
  \frac{m_{E,M}^d}{T} = a_{E,M}^d\Bigg[a+
  c_{E,M;1}^d\frac{|e|B}{T^2}\mathrm{atan}
  \Bigg(\frac{c_{E,M;2}^d}{c_{E,M;1}^d}\frac{|e|B}{T^2}\Bigg)\Bigg]
\end{equation}
where $d$ indicates the spatial direction and the parameters
$a^d_{E,M}$, $c_{E,M;1}^d$ and $c_{E,M;2}^d$ are determined by the
regression procedure. The way we wrote the ansatz above makes the
interpretation of these parameters simple: the constant $a$ represents
the ratio $m/T$ at $B=0$, while the $c_1$ and $c_2$ coefficients are,
respectively, the asymptotic linear slope and the quadratic constant
describing the behaviour of the masses as a function of
$|e|B/T^2$. The results of the regression are reported in
Tab.~\ref{tab:fitparams}, while the best fit curves are shown together
with numerical data in Fig.~\ref{fig:masses}. As can be seen, the fit
results confirm what pointed out before: the large $B$ slope of the
two masses is compatible and, in the color-magnetic case, an
anisotropy emerges. At the same time, noise affects the color-electric
masses and the small magnetic field regime, so as to reduce the
precision of the parameters of the model.

\begin{table}[thb]
  \small
  \centering
  \caption{Regression parameter obtained fitting the model in
    Eq.~\eqref{eq:ansatz}.}
  \label{tab:fitparams}
  \begin{tabular}{lllll}\toprule
    ~ & $a$ &  $c_1$ &  $c_2$  & $\chi^2/\mathrm{d.o.f}$\\\midrule
    $m^{XY}_{M}$  & 4.964(82) & 0.137(19)$\times 10^{-1}$ & 0.141(55)$\times 10^{-2}$  &1.06 \\
    $m^{Z~~}_{M}$ & 4.935(79) & 0.099(20)$\times 10^{-1}$ & 0.094(49)$\times 10^{-2}$  &1.10 \\
    $m^{XY}_{E}$  & 9.24(21)  & 0.120(47)$\times 10^{-1}$ & 0.069(38)$\times 10^{-2}$  &0.63 \\
    $m^{Z~~}_{E}$ & 9.34(20)  & 0.17(28) $\times 10^{-1}$ & 0.039(21)$\times 10^{-2}$  &0.85 \\\bottomrule
  \end{tabular}
\end{table}

\section{Conclusions}\label{conclusions}
We have shown that the screening masses in QCD are deeply modified by
the presence of a strong external magnetic field. Our results,
obtained on the lattice at a single lattice spacing, agree previous
determinations of the masses and suggest that they grow almost
linearly in $B$ keeping the mass hierarchy. Anisotropies emerge
expecially in the magnetic mass, while in the electric case the effect
may be hindered by the large relative errors. Magnetic effets are reduced
when the tempereture increase and the essential properties of the
masses are well described in terms of the dimensionless ratio
$|e|B/T^2$.

Our data agree qualitatively the results expected by perturbative
calculations \cite{Alexandre:2000jc,Bandyopadhyay:2016fyd}. In
addition, the results we obtained resembles the behaviour observed in
the finite temperature regime below the pseudo-critical temperature
$T_c$ \cite{Bonati:2016kxj}. Indeed, in the deconfined region the
magnetic field is responsable of a early suppression of the confining
properties while it enhances the screening effect in the deconfined
regime. The main clue comes from the intepretation of a decreasing
$T_c$ as a function of $B$ \cite{Bali:2011qj}, so that at low
temperatures the system approaches the transition earlier, while gets
farther from it above. Finally, as also pointed out initially in the
introduction to our study, the modifications of the screening masses
may have a relevant role in heavy-ion collisions, expecially in the
producion of heavy mesons. However, quantitative predictions may be
done only carrying on an in-depth investigation of the features of the
magnetic fields produced in the collisions and of their effect on the
bound states.

\section*{\small Acknowledgments}
\small We acknowledge PRACE for giving us access to resource FERMI at
CINECA in Italy, under project ${\text{Pra09-2400 - SISMAF}}$.

\clearpage
% \bibliography{lattice2017}

%%%%%%%%%%%%%%%%%%%%%%%%%%%%%%%%%%%%%%%%%%%%%%%%%%%%%%%%%%%%%%%%%%%%%%%%%%%%%
\end{document}